# Optimal dimensions of cone and pyramid moth-eye structures for SiO$_2$ windows

**Chaoran Tu[1,*], Jonathan Hu[2], Curtis R. Menyuk[1], Thomas F. Carruthers[1], L. Brandon Shaw[3], Lynda E. Busse[3], and Jasbinder S. Sanghera[3]**

[1]*Department of Computer Science and Electrical Engineering, University of Maryland Baltimore County, Baltimore, MD 21250, USA*
[2]*Department of Electrical and Computer Engineering, Baylor University, Waco, TX 76798, USA*
[3]*US Naval Research Lab, 4555 Overlook Ave SW, Washington, DC 20375, USA*
*chaoran1@umbc.edu

**Abstract:** We computationally investigate the transmission efficiency through moth-eye nanostructures that are fabricated on SiO$_2$ windows in the wavelength range from 0.4 to 2 $\mu$m. We investigated both truncated cones and truncated pyramids, and we varied the height, bottom width, and top width of these shapes in order to maximize the transmission efficiency. We found that there is no substantial difference in transmission between truncated cone and pyramid structures. Using the constraints from the current achievable experimental limits, a relatively uniform transmission coefficient of larger than 98.8% can be obtained from 0.4 $\mu$m to 2 $\mu$m. These transmission results are only 0.4% in absolute value lower than the transmission of a structure that is not constrained by current experimental limits.

## 1. Introduction

Moth-eye structures are made of a biomimetic and periodic array of nanostructures on a substrate, which can reduce the Fresnel reflections by reducing the impedance mismatch of the air and the substrate [1–5]. The moth-eye structures introduce a smooth transition between one medium and another, ensuring that incident light does not encounter a sudden change in the refractive index, which would cause a strong reflection. Similar to conventional multilayer antireflection (AR) coatings [6–9], moth-eye structures create a gradually varying effective refractive index profile that is determined by the shapes of the moth-eye elements and their arrangement on the surface. Moth-eye structures have several important advantages over traditional thin-film AR coatings, including single material fabrication, minimal surface preparation, environmental tolerance, surface adhesion, and self-cleaning via the lotus effect [10,11]. Additionally, it has also been shown that in many cases periodic moth-eye structures have a higher laser-induced damage threshold than do traditional AR-coated surfaces [12–16]. Moth-eye structures are currently used in many applications, including flexible display devices [17], automotive glass [10,18,19], laser systems [20–23], fiber optics [20], and photovoltaics [24–29].

In prior work, Busse et al. [15] fabricated fused silica glass windows with optical transmission greater than 99.5% for wavelengths between 0.775 $\mu$m and 1.35 $\mu$m, and we have achieved a theoretical transmission larger than 99.5% over the wavelength range from 0.5 $\mu$m to 2.0 $\mu$m using moth-eye cone structures [30]. We note that the shapes of the moth-eye structures are generally cones or pyramids [10] and an exploration of the theoretical performance limit of the pyramid structure has not been carried out to date. In this work, we compare the moth-eye structures that are composed of either truncated cones or truncated pyramids, and we vary their heights, top widths, and bottom widths to maximize the transmission. We computationally investigate the transmission efficiency through one-layer of moth-eye nanostructures that are fabricated on SiO$_2$ windows in the wavelength range from 0.4 to 2 $\mu$m.

The remainder of this paper is organized as follows: In Sec. 2, we introduce our computational model and the finite-difference time domain (FDTD) software that we use. In Sec. 3, we show our optimization results. In Sec. 4, we compare the optimal spectra. Finally, we conclude in Sec. 5.

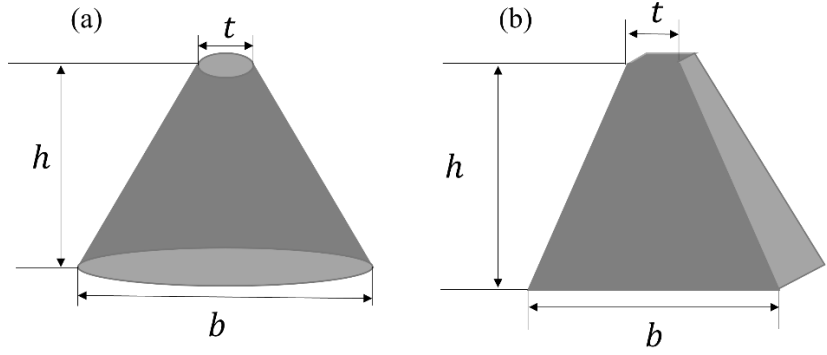

Fig. 1. Different shapes used in the moth-eye structure: (a) the truncated cone and (b) the truncated pyramid.

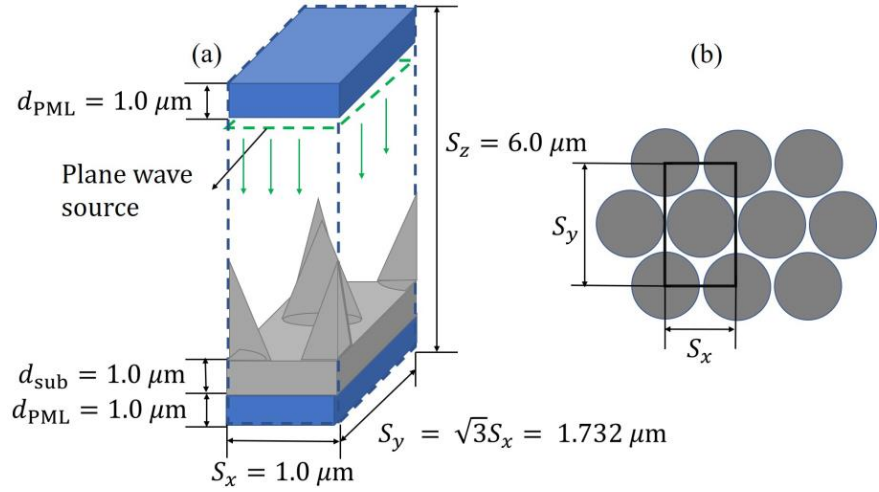

Fig. 2. Schematic view of silica moth-eye structure: (a) A three-dimensional view of the computational unit cell. (b) A top view showing a unit cell.

## 2. Computational model

Figure 1 shows a schematic illustration of truncated cones and truncated pyramids. The bottom width, top width, and height are denoted by $b$, $t$, and $h$, respectively. Due to experimental limits, we focused on moth-eye structures with a minimum top width of 0.15 $\mu$m and a maximum height of 0.6 $\mu$m. However, to explore the theoretically obtainable maximum transmissions, we also considered structures with larger heights and smaller top widths. In Fig. 2, we show a schematic illustration of our computational model. In our simulation, we calculated the transmission of a normally incident plane wave through the $SiO_2$ nanostructured surface using the open-source FDTD software MEEP that is available under the GNU General Public License [26]. We used a minimum spatial resolution of 5 nm in our simulations, and we verified the convergence of the results in all cases. We took advantage of the hexagonal symmetry of the structure and periodicity of the optical field to use the computational grid that we show in Fig. 2, which consists of a single unit cell in the $x$- and $y$-directions. The unit cell has dimensions $S_x =$ 1.0 $\mu$m, $S_y = \sqrt{3}S_x = 1.732$ $\mu$m. We used 1 $\mu$m perfectly matched layers (PMLs) ($d_{PML} = 1.0$ $\mu$m) at the top and the bottom [26]. We used a broadband plane wave source that is located just below the upper PML. We calculated transmission and reflection spectra by taking a Fourier transform of the time-domain flux through surfaces lying just below the moth-eye structure and above the plane wave source.

## 3. Results

We used $h = 0.6$ $\mu$m, $b = 0.95$ $\mu$m, and $t = 0.15$ $\mu$m as our baseline parameters. The values of $h$ and $t$ correspond to the achievable experimental limits. In Fig. 3, we compare the transmission spectra for two orthogonal polarizations with these baseline parameters. The polarization angle in Fig. 3 refers to the angle between the electric field direction and the $x$-axis. The moth-eye structures with cones are strictly hexagonally symmetric, so that the transmission has no polarization dependence [30]. By contrast, the square cross-section of the pyramids breaks the symmetry of the structures and introduces a small but negligible polarization dependence. For simplicity, the remainder of our results are all obtained with a polarization angle of 0º. We see that all curves show sharp dips that we attribute to resonances with the doubly periodic structure. Similar resonant dips appear in all cases as we modify the structure parameters, but the location of the resonances depends sensitively on these parameters. In an experimental setting, where the structure parameters cannot be precisely controlled, it seems likely that these sharp dips would be smoothed out.

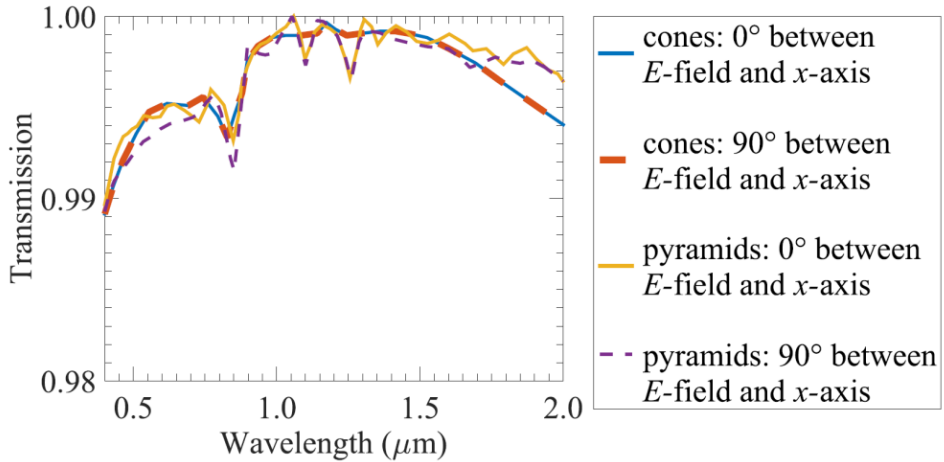

.

Fig. 3. Transmission spectra of truncated cones and pyramids with normally incident light and with polarization angles of 0º and 90º.

*3.1 Bottom width*

In Fig. 4, we show the transmission spectra for different values of the bottom width $b$ for both truncated cones and pyramids. We considered the truncated cones, and we allowed $b$ to vary from 0.15 $\mu$m to 0.95 $\mu$m, which is within the range that can be fabricated. Next, we considered truncated pyramids, and we allowed $b$ to vary from 0.15 $\mu$m to 0.8 $\mu$m.

For the truncated cone structures, Fig. 4(a) shows that the transmission increases in the entire wavelength range by about 0.03 when the bottom width $b$ of the truncated cones increases from 0.15 $\mu$m to 0.95 $\mu$m. For the truncated pyramid structures, Fig. 4(b) shows that the transmission generally increases in the entire wavelength range by about 0.03 when the bottom width $b$ of the truncated pyramids increases from 0.15 $\mu$m to 0.8 $\mu$m. We observe that the results are similar for truncated cone and pyramid structures.

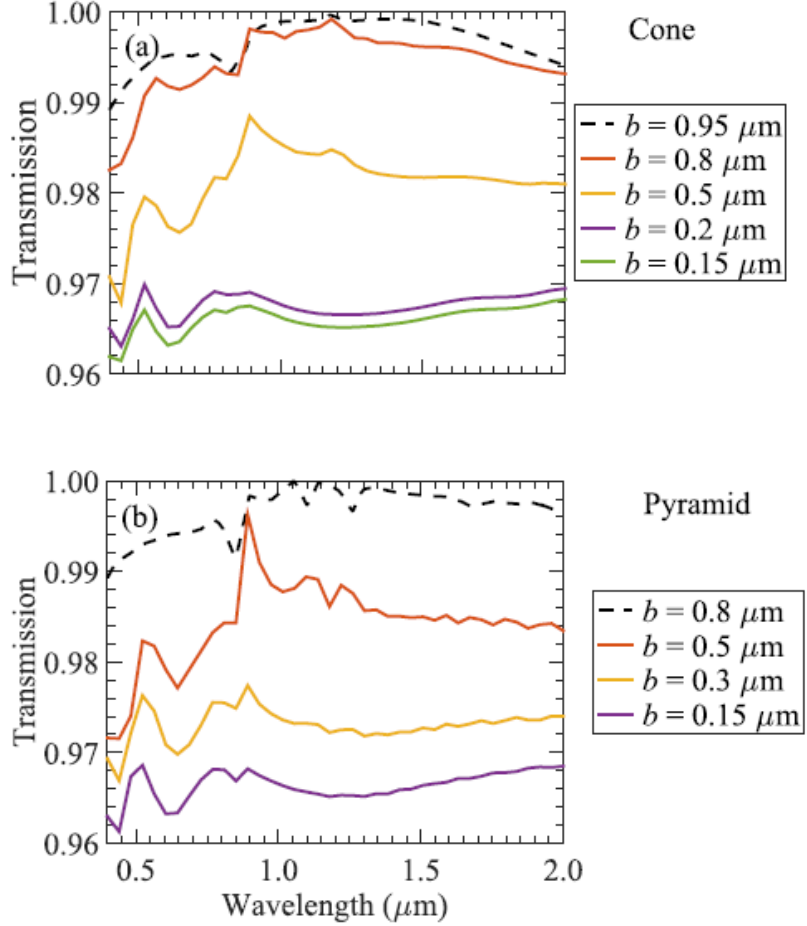

Fig. 4. Transmission spectra when varying the bottom width $b$ of (a) the truncated cones ($t = 0.15$ $\mu$m and $h = 0.6$ $\mu$m) and (b) the truncated pyramids ($t = 0.15$ $\mu$m and $h = 0.6$ $\mu$m). The black dashed line corresponds to the currently achievable experimental limit of $b = 0.95$ $\mu$m for cone structures and $b = 0.8$ $\mu$m for pyramid structures.

*3.2 Top width*

In Fig. 5, we show the transmission spectra for different values of the top width $t$ with truncated cones and pyramids. For the truncated cone structures, Fig. 5(a) shows that the transmission spectrum decreases in the entire wavelength range by about 0.02, when the top width $t$ of the truncated cones increases from 0 to 0.95 $\mu$m. We also find that if we reduce the top of the cone to 0.15 $\mu$m, we almost reach the theoretically optimal transmission spectrum with $t = 0$. For the truncated pyramid structures, Fig. 5(b) shows that the transmission spectrum decreases in the entire wavelength range by about 0.02 when the top width $t$ of the truncated pyramids increases from 0 to 0.8 $\mu$m. We also find that structures with $t = 0.15$ $\mu$m have a transmission spectrum that is close to the optimal transmission spectrum that is achieved for the top width $t = 0$ $\mu$m for the pyramids.

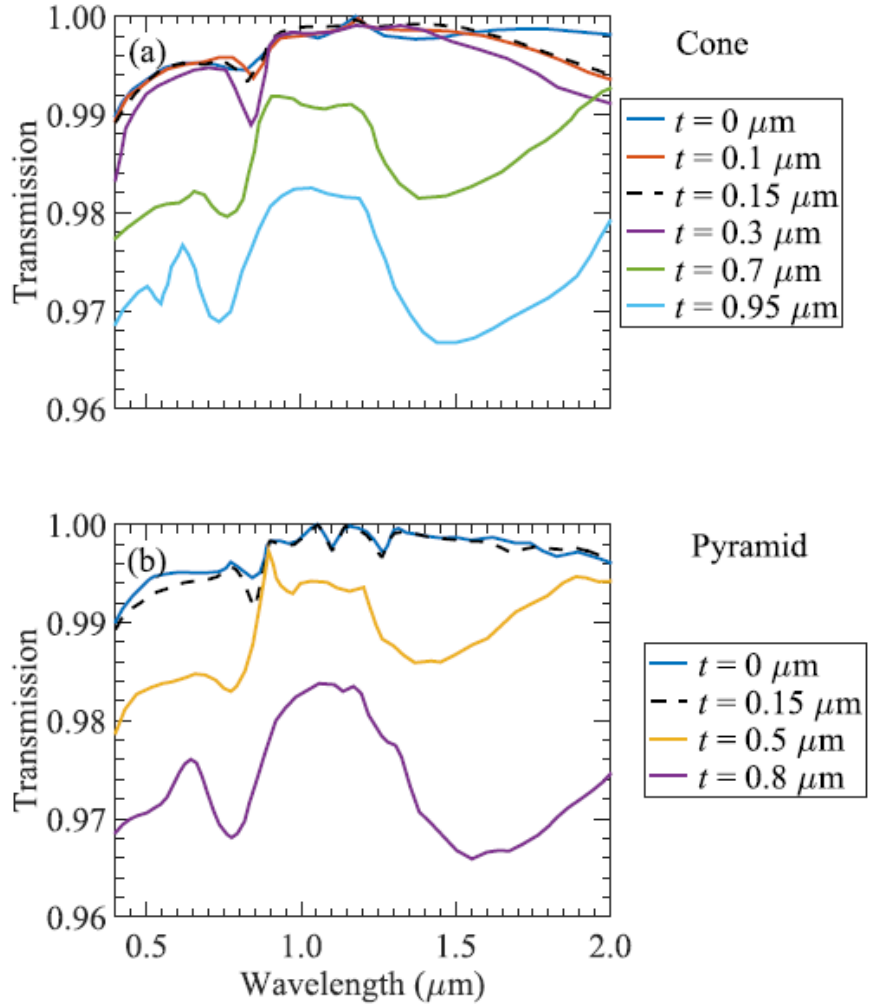

Fig. 5. Transmission spectra when varying the top width $t$ of (a) the truncated cones ($b = 0.95\ \mu$m and $h = 0.6\ \mu$m) and (b) the truncated pyramids ($b = 0.8\ \mu$m and $h = 0.6\ \mu$m). The black dashed line in each plot corresponds to the currently achievable experimental limit of $t = 0.15\ \mu$m.

*3.3 Height*
In Fig. 6, we show the transmission spectra for different values of the height $h$ with truncated cones and pyramids. For both structures, we allowed $h$ to vary between 0.2 $\mu$m and 1.6 $\mu$m. To determine the theoretically achievable optimum, we also allowed $h$ to be larger than 1.6 $\mu$m. For the truncated cone structures, Fig. 6(a) shows that the transmission spectrum generally increases, when the height $h$ of the truncated cones increases from 0.2 $\mu$m to 1.6 $\mu$m. Beyond $h = 0.6\ \mu$m, there is no significant overall improvement in the transmission spectrum. For the truncated pyramid structures, Fig. 6(b) shows that the transmission spectrum generally increases when the height $h$ of the truncated pyramids increases from 0.2 $\mu$m to 1.6 $\mu$m. There is no significant improvement over the entire wavelength range beyond $h = 0.6\ \mu$m.

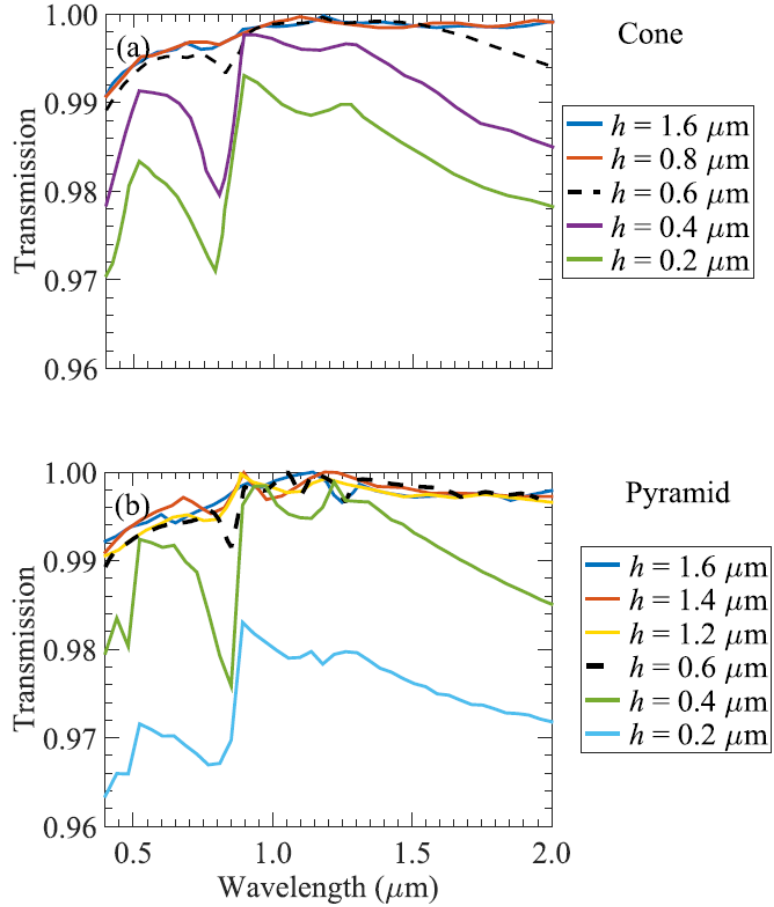

Fig. 6. Transmission spectra when varying the height $h$ of (a) the truncated cones ($t = 0.15\ \mu$m and $b = 0.95\ \mu$m) and (b) the truncated pyramids ($t = 0.15\ \mu$m and $b = 0.8\ \mu$m). The black dashed line corresponds to the currently achievable experimental limit.

## 4. Comparison of the optimal spectra

Based on the transmission spectra, we conclude that a truncated cone moth-eye structure with a top width $t$ that is as small as possible, a bottom width $b$ that is as large as possible, and a cone height $h$ that is as high as possible will have a high transmission spectrum over a wide wavelength range. In Fig. 7, we compare the transmission spectra for both truncated cone and pyramid structures. The optimal dimensions of the truncated cone and pyramids that are consistent with the current experimental limits are: $t = 0.15\ \mu$m, $b = 0.95\ \mu$m, $h = 0.6\ \mu$m and $t = 0.15\ \mu$m, $b = 0.8\ \mu$m, $h = 0.6\ \mu$m, respectively. The minimum transmission is 98.8% for our entire wavelength range of interest from 0.4 to 2 $\mu$m. If we allowed the parameters to exceed the experimental limits, then the optimal parameters are $t = 0\ \mu$m, $b = 0.95\ \mu$m, $h = 1.6\ \mu$m for truncated cone structures and $t = 0\ \mu$m, $b = 0.8\ \mu$m, $h = 2.2\ \mu$m for truncated pyramid structures. The minimum transmission is 99.2% for our entire wavelength of interest. We observe that remaining within the experimental limits leads to a transmission spectrum of more than 98.8% over our entire wavelength range of interest, yielding a penalty in the transmission spectrum of less than 0.4% in an absolute value, compared to an ideal structure. We also find that there is no substantial difference in transmission between truncated cone and pyramid structures.

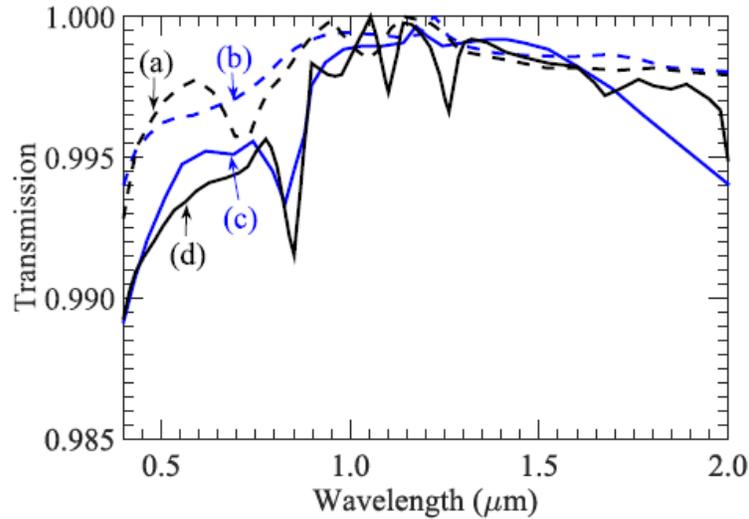

Fig. 7. Comparison of the transmission spectra of (a) truncated pyramids with the theoretically optimal parameters ($t = 0$ μm, $b = 0.8$ μm and $h = 2.2$ μm), (b) truncated cones with the theoretically optimal parameters ($t = 0$ μm, $b = 0.95$ μm and $h = 1.6$ μm), (c) truncated cones with the experimentally-limited optimal parameters ($t = 0.15$ μm, $b = 0.95$ μm and $h = 0.6$ μm), and (d) truncated pyramids with the experimentally-limited optimal parameters ($t = 0.15$ μm, $b = 0.8$ μm and $h = 0.6$ μm).

## 5. Conclusion

We used the FDTD method to computationally study the transmission of light that is normally incident on a $SiO_2$ glass window with moth-eye structures that use truncated cones and pyramids. We investigated the effect of changing the top width, bottom width, and height of truncated cones and pyramids. The transmission generally increases when truncated cones and pyramids have a narrow top width, a wide bottom width, and a large height. We find that there is no substantial difference in transmission between truncated cone and pyramid structures. Within the current achievable experimental fabrication limits ($t = 0.15$ μm, $h = 0.6$ μm with $b = 0.95$ μm for truncated cone structures and $b = 0.8$ μm for truncated pyramid structures), the optimal truncated cone and pyramid moth-eye structures have a relatively uniform transmission coefficient, which is larger than 98.8% from 0.4 μm to 2 μm. Using the constraints from the current achievable experimental limits, the minimum transmission is only 0.4% (in absolute value) lower than the minimum transmission of 99.2% for the theoretically optimal moth-eye structures for both truncated cones and pyramids. Although the optimal structures can give slightly higher transmission, the current fabrication limitations already give promising results close to the optimal structures.